\documentclass[intlimits,twoside,a4paper]{article}

\usepackage{amsmath,amssymb}
\usepackage{graphicx}
\usepackage{wrapfig}

\usepackage[T2A]{fontenc}
\usepackage[cp1251]{inputenc}
\usepackage[eqsecnum]{cmpj2}
\issue{2012}{15}{3}{33002}
\doinumber{10.5488/CMP.15.33002}

\providecommand{\abs}[1]{\lvert#1\rvert}
\DeclareMathOperator{\Sp}{Sp}
\allowdisplaybreaks[3]

\title[Two-state Bose-Hubbard model in HCB limit: Non-ergodicity]%
{The two-state Bose-Hubbard model in the hard-core boson limit:
Non-ergodicity and the Bose-Einstein condensation}
\author{I.V. Stasyuk, O.V. Velychko}

\authorcopyright{I.V. Stasyuk, O.V. Velychko, 2012}
\address{Institute for Condensed Matter Physics
of the National Academy of Sciences of Ukraine, \\1~Svientsitskii
Str., 79011 Lviv, Ukraine}

\date{Received July 11, 2012, in final form July 13, 2012}
\begin{document}

\maketitle

\begin{abstract}
The Bose-Einstein condensation in the hard-core boson limit (HCB) of the
Bose-Hubbard model with two local states and the particle hopping in the
excited band only is investigated. For the purpose of  considering the
non-ergodicity,  a single-particle spectral density is calculated in the
random phase approximation by means of the temperature boson Green functions.
The non-ergodic contribution to the momentum distribution function of
particles (connected with the static density fluctuations) increases
significantly and becomes comparable with the ergodic contribution   in the superfluid
phase near the tricritical point.
\pacs 03.75.Hh, 03.75.Lm, 64.70.Tg, 71.35.Lk, 37.10.Jk, 67.85.-d
\end{abstract}

%\renewcommand\encodingdefault{T2A}

%\selectlanguage{ukrainian}

\section{Introduction}

The Bose-Hubbard model (BHM) \cite{Fisher89} is widely used in describing the
thermodynamics and the dynamics of Bose atoms in optical lattices
\cite{Greiner02a,Greiner02b}. Such systems demonstrate the phase transition
into the phase with Bose-Einstein (BE) condensate at very low temperatures.
Thus, the system can be in the normal (NO) phase (the state of the so-called
Mott insulator, MI), or in the phase with the BE condensate (the superfluid
state, SF). The BHM can be used for description of various phenomena such as the quantum delocalization of hydrogen atoms adsorbed on the
surface of transition metals \cite{Nishijima05}, quantum diffusion of light
particles on the surface or in the bulk \cite{Reilly91}, thermodynamics of
the impurity ion intercalation into semiconductors \cite{Mysakovych10}.

The BHM has been intensively studied for the two recent decades starting with the
pioneering work \cite{Sheshadri93}, where the thermodynamics of the model was
considered in the mean field approximation (MFA) for a particle hopping
with an exact treatment of the Hubbard repulsion. Such a description revealed
the first order NO-SF phase transition for particles residing in the ground
state of local wells.
A prospective generalization of the BHM consists in taking into account the
excited vibrational states of bosons in local potential minima of lattice
sites. For example, intersite particle hopping through the excited states is
much easier in the case of  quantum delocalization or diffusion
\cite{Puska85,Brenig93} as well as in the case of optical lattices \cite{Mueller07}.
However, a possible BE condensation in the excited states was considered for the case of optical pumping in order to  increase  their occupations
\cite{Isacsson05}. In this case, the orbital degeneration of the excited
$p$-state can lead to the appearance of a special type of the BE condensate.

A thorough study of the phase transition into the phase with BE condensate in
the BHM with two local states and the boson hopping only in the excited band
was carried out  in works \cite{Stasyuk11CMP,Stasyuk11TMF}. In the MFA and the
hard-core boson (HCB) limit, the instability related to the NO-SF transition
was studied which occurs at excitation energies lower than the absolute value
of the hopping parameter. Conditions of the change of the  above mentioned
phase transition from the second order to the first one were studied. The
respective phase diagrams $(\Theta,\mu)$ and $(\abs{t_0'},\mu)$ were
analyzed. A possible phase separation into NO and SF phases at a fixed
concentration was revealed. Two-time boson Green's functions (GF) and a
single-particle spectral density were calculated in the random phase
approximation (RPA). The structure of excitation spectra of the particle
and hole types in the region of the NO-SF phase transition was
investigated.

Some physical many-particle systems do not hold the ergodicity hypotesis. It
means that there are regions of the phase space of a system which cannot
be reached by the trajectory of the system evolution. From practical
viewpoint, non-ergodicity leads to a distinction between isothermal
and  isolated responses (susceptibilities) of a system caused by the
zero-frequency term present in the isothermal response only. In this case, the use of
the  ``isolated'' (Kubo) response leads to  erroneous values of such
physical characteristics as compressibility, specific heat and susceptibility
for the systems described by means of the Ising model
\cite{Baryakhtar84eng,Izyumov88eng} (a zero Kubo response), the
pseudospin-electron model \cite{Stasyuk94CMP} and the Falicov-Kimball model
(the non-ergodic contribution describes the singularity related to the phase
transition \cite{Shvaika00,Shvaika01}). The BHM in the HCB limit, as will be
shown below, also belongs to the non-ergodic class. The mentioned problems
can be solved  using the temperature (Matsubara's) GF
\cite{Matsubara55} (see, e.g., a scheme for calculation of many-time
correlation functions \cite{Shvaika06}).

In the present article we study the non-ergodic contribution to the particle
momentum distribution function by the example of the two-state Bose-Hubbard
model in the HCB limit. A detailed investigation is performed in the region
close to the tricritical point where the order of the phase transition to the
SF phase changes to the first one.

\section{Thermodynamics of the model in the hard-core boson limit}

The Hamiltonian of the BHM takes into account the tunnel hopping of particles
between the nearest neighbour lattice sites and the mutual repulsion of the
particles located in the same potential well \cite{Fisher89}:
\begin{equation}
    \hat{H} =
    \sum_{ij} t_{ij} b_i^+ b_j
    + \frac{U}{2}\sum_i n_i(n_i-1)
    - \mu\sum_i n_i
    \,.
    \label{eq2-01}
\end{equation}
Here, $t_{ij}$ characterizes the hopping energy, $U$ is the energy of the
Hubbard intrasite interaction ($U>0$), $\mu$ is the chemical potential of the
boson particles, $b_i^+, b_j$ are the Bose operators of particle creation and
annihilation on the $i$-th site in the ground vibrational state. Taking into
account the first excited state and considering the particle hopping only
between these excited states, one can generalize the Hamiltonian
\eqref{eq2-01} as follows:
\begin{align}
    \hat{H} &= (\varepsilon-\mu)\sum_i b_i^+ b_i
        + (\varepsilon'-\mu)\sum_i c_i^+ c_i
        + \frac{U_b}{2}\sum_i n_i^b(n_i^b-1)
    \notag\\
    &\quad
        + \frac{U_c}{2}\sum_i n_i^c(n_i^c-1)
        + U_{bc}\sum_i n_i^b n_i^c
        + \sum_{ij} t_{ij}' c_i^+ c_j\,,
    \label{eq2-02}
\end{align}
where $c_i^+, c_i$ are the Bose operators of the particles in the excited
state, $\varepsilon$ ($\varepsilon'$) is the energy of the particle in the
ground (excited) state, $U_b,U_c,U_{bc}$ are the parameters of the Hubbard
repulsion.

Starting from the basis of the single-site states $|i;n_i^b,n_i^c\rangle$,
formed by the particle occupation numbers (eigenvalues of the operators
$n_i^b=b_i^+ b_i$ and $n_i^c=c_i^+ c_i$), one can introduce the Hubbard
operators \cite{Hubbard65IV}
\begin{equation}
    X_i^{n,m;n',m'} \equiv |i;n,m\rangle\langle i;n',m'|.
    \label{eq2-03}
\end{equation}

Let us restrict our study to the HCB case (when $n+m\leqslant1$). In the
considered model, such a situation is reached at $U_b,U_c,U_{bc}\to\infty$.
In this limit, the single-site Hamiltonian becomes three-level with the
respective energies
    $\lambda_{0}=0$,
    $\lambda_{1}=-\mu$,
    $\lambda_{2}=\delta-\mu$
(the following shortened designations for single-site states are used:
$| 0 \rangle \equiv | 00 \rangle$,
$| 1 \rangle \equiv | 10 \rangle$,
$| 2 \rangle \equiv | 01 \rangle$).
Here $\delta=\varepsilon'-\varepsilon$ is the energy of transition to the
local excited vibrational state. Respectively,
\begin{align*}
     b_i  &=  X_i^{00,10}  \equiv  X_i^{01},&
     c_i  &=  X_i^{00,01}  \equiv  X_i^{02};
     \\
     n^b_i  &=  X_i^{10,10}  \equiv  X_i^{11},&
     n^c_i  &=  X_i^{01,01}  \equiv  X_i^{22}.
\end{align*}
Finally, the Hamiltonian \eqref{eq2-02} is formulated in the $X$-operator
representation
\begin{equation}
    \hat{H} = \sum_{ip} \lambda_p X_i^{pp}
    + \sum_{ij} t_{ij}' X_i^{20} X_j^{02}.
    \label{eq2-09}
\end{equation}

The BE condensation takes place in the band that originated from the hopping
between the excited states of the neighbouring wells (see below). Thus, the order
parameter is equal to the average of boson creation or annihilation operators
    $\xi  = \langle X_i^{20} \rangle = \langle X_i^{02} \rangle$\;
    $(\xi = \langle c_i^+ \rangle = \langle c_i \rangle)$.

Let us separate a MFA part of the Hamiltonian \eqref{eq2-09}
\begin{equation}
    \hat{H}_{\mathrm{MF}} =
    -N t_0'\xi^2+\sum_{ip} \lambda_p X_i^{pp}
    +t_0'\xi\sum_{i}  (X_i^{20}+X_i^{02}),
    \label{eq2-11}
\end{equation}
where $t_0'$ is the Fourier transform of the hopping energy $t_{ij}'$ at
$\vec{q}=0$ (hereinafter we consider $t_0'<0$, i.e., assuming symmetrical
excited states, while $t_0'>0$ for antisymmetrical ones \cite{Liu06}).
The $\xi$ parameter is obtained from the self-consistency equation
$\xi {=} Z^{-1}\Sp[X_i^{20}\*\exp(-\beta\hat{H}_{\mathrm{MF}})]$,
where $Z=\Sp\exp(-\beta\hat{H}_{\mathrm{MF}})$.

This approximation corresponds to the exact solution for the case of an
infinite-range hopping. A strict derivation of the above statement for the
standard BHM has been given in the work \cite{Bru03} (see also
\cite{Dorlas06}) based on the approach in the style of the
Bogolyubov-Tyablikov variational method.

Finally, a complete Hamiltonian looks as follows:
\begin{equation}
    \hat{H} =
    \hat{H}_{\mathrm{MF}}
    +\sum_{ij}t_{ij}'(X_i^{20}-\xi)(X_j^{02}-\xi).
    \label{eq2-13}
\end{equation}

The MF Hamiltonian $\hat{H}_{\mathrm{MF}}$ can be reduced to a diagonal
form by the transformation
\begin{equation}
    \left(
    \begin{array}{l}
        | 0 \rangle \\
        | 1 \rangle \\
        | 2 \rangle
    \end{array}
    \right)
    =
    \left(
    \begin{array}{ccc}
        \cos\vartheta & 0 & -\sin\vartheta \\
        0 & 1 & 0 \\
        \sin\vartheta & 0 & \cos\vartheta
    \end{array}
    \right)
    \left(
    \begin{array}{l}
        | \tilde{0} \rangle \\
        | \tilde{1} \rangle \\
        | \tilde{2} \rangle
    \end{array}
    \right),
    \label{eq2-14}
\end{equation}
where
\begin{equation}
    \cos2\vartheta = (\delta-\mu)/E,
    \qquad
    \sin2\vartheta = 2\abs{t_0'}\xi/E;
    \qquad
    E=\sqrt{(\delta-\mu)^2+4(t_0'\xi)^2}.
    \label{eq2-15}
\end{equation}
In terms of operators
$\widetilde{X}^{\mu\nu}=|\tilde{\mu}\rangle\langle\tilde{\nu}|$
\begin{equation}
    \hat{H}_{\mathrm{MF}} =
    -Nt_0'\xi^2
    +\sum_{i\mu}\tilde{\lambda}_\mu\widetilde{X}_i^{\mu\mu}.
    \label{eq2-16}
\end{equation}
Energies of single-site states $\tilde{\lambda}_\mu$ in the phase with a
broken symmetry ($\xi\neq0$) are equal to
    $\tilde{\lambda}_{0,2} = \frac{1}{2}(\delta-\mu \mp E)$,
    $\tilde{\lambda}_1 = -\mu$.

It is convenient to introduce the following linear combinations
\begin{equation}
    \textstyle
    \sigma_i^z =
        \frac{1}{2}(\widetilde{X}_i^{00}-\widetilde{X}_i^{22}),
    \qquad
    \sigma_i^+ \equiv \sigma_i^x+\mathrm{i}\sigma_i^y
        =\widetilde{X}_i^{02},
    \qquad
    \sigma_i^- \equiv \sigma_i^x-\mathrm{i}\sigma_i^y
        =\widetilde{X}_i^{20}.
    \label{eq2-19}
\end{equation}
Operators $\sigma_i^{\alpha}$ have the same properties as the spin operators
for the case $S=1/2$ and fulfil the same commutation rules (with an important
difference: the anticommutator of $\sigma_i^+$ and $\sigma_i^-$ operators
is equal to $\widetilde{X}_i^{00}+\widetilde{X}_i^{22}$ instead of unity as in
the ordinary case).

Thus, the $X$-operators are represented on a new basis as follows:
\begin{equation}
    X_i^{20} =
    \sin2\vartheta\,\sigma_i^z
    +\cos2\vartheta\,\sigma_i^x
    -\mathrm{i}\sigma_i^y\,,
    \qquad
    X_i^{02} = (X_i^{20})^{+},
    \label{eq2-20}
\end{equation}
with the MF Hamiltonian in the form
\begin{equation}
    \hat{H}_{\mathrm{MF}} =
    N
    \left(
    \frac{\delta-\mu}{2}-t_0'\xi^2
    \right)
    -\frac{\delta+\mu}{2}\sum_{i}\widetilde{X}_i^{11}
    -E\sum_{i}\sigma_i^z\,.
    \label{eq2-21}
\end{equation}

\begin{wrapfigure}{i}{0.5\textwidth}
%\begin{figure}%[!b]
%
\centerline{\includegraphics[width=0.46\textwidth]{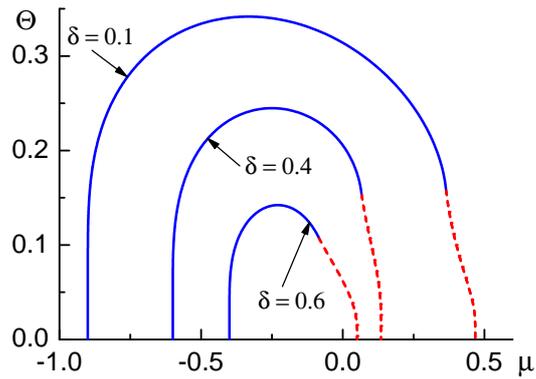}}%
\caption{Phase diagrams $(\Theta,\mu)$ at various values of $\delta$
($\abs{t_0'}=1$). The solid line corresponds to the second order phase
transition and the dashed line denotes the first order transition.}
\label{fig03}
\vspace{8mm}
%\end{figure}
\end{wrapfigure}
The Hamiltonian \eqref{eq2-13} expressed in terms of operators \eqref{eq2-19}
is a three-state generalization of the standard Hamiltonian of the HCB model
\cite{Micnas90}. The presence of the third level $|\tilde{1}\rangle$ does not
effect the pseudospin dynamics but modifies the occupations of levels
$|\tilde{0}\rangle$ and $|\tilde{2}\rangle$, changing this way the
thermodynamic behaviour of the model. Moreover, the application of pseudospin
formalism allows one to utilize the usual scheme of the random phase
approximation (RPA) for Green's functions of spin models, originated from the
well-known Tyablikov decoupling for the Heisenberg model
\cite{Bogolyubov59eng}.

Let us make a short revision of the thermodynamics of a three-state model in
the MFA. Using the Gibbs distribution with the Hamiltonian \eqref{eq2-16} and
taking into account that the average values of transverse pseudospin
projections $\langle \sigma_i^{\pm} \rangle$ are equal to zero, one can
obtain the equation for the order parameter
$\xi=\langle X_i^{02} \rangle$:
\begin{equation}
    \xi =
    Z^{-1}E^{-1}\abs{t_0'}\xi
    \left(
    \mathrm{e}^{-\beta\tilde{\lambda}_0}
    -
    \mathrm{e}^{-\beta\tilde{\lambda}_2}
    \right)
    .
    \label{eq3-02}
\end{equation}
The case of $\xi=0$ corresponds to the NO phase. One can obtain nonzero solutions
(indicating the appearance of the BE condensate)  from the following
equation
\begin{equation}
    Z^{-1}E^{-1}\abs{t_0'}
    \left(
    \mathrm{e}^{-\beta\tilde{\lambda}_0}
    -
    \mathrm{e}^{-\beta\tilde{\lambda}_2}
    \right)
    = 1
    .
    \label{eq3-03}
\end{equation}

It has been shown in work \cite{Stasyuk11CMP} that in the region $\mu<0$,
nonzero values of $\xi$ gradually appear at the second order transition while
in the region $\mu>0$ at a low enough temperature, the curve $\xi(\mu)$ is
S-shaped which indicates the first order phase transition. The line of this
transition is defined by the equality of values of the grand canonical
potentials $\Omega=-\Theta\ln Z$ in phases NO and SF. The change of the phase
transition order and the localization of the respective tricritical points
are presented in the phase diagrams in figure~\ref{fig03}.

Unlike the standard (two-level) HCB model (where  the Bose-particles remain in
the local ground state, the excited local states are neglected and phase
transitions between NO and SF phases are always of the second order), the
phase diagrams in figure~\ref{fig03} are asymmetrical.

\section{Temperature Green's functions and the non-ergodic contribution}

The spectrum of boson particles and the single-site spectral
density for this model are calculated in the RPA using the
two-time GF in work \cite{Stasyuk11TMF}. Herein below we will perform
similar calculations in the RPA but applying the temperature GF.

Assuming the hopping between the excited states only, the dynamics of
the model corresponds to the HCB model and we consider only GF constructed on
the operators which describe the transition between the states
$\lvert{0}\rangle$ and $\lvert{2}\rangle$
\begin{align}
    \left\langle
        T_{\tau}X_i^{20}(\tau)X_j^{02}(\tau')
    \right\rangle
    &=
    \xi^2
    +
    \left\langle
        T_{\tau}(X_i^{20}-\xi)_{\tau}(X_j^{02}-\xi)_{\tau'}
    \right\rangle
    ,
    \label{eqii-01}
    \notag
    \\
    \left\langle
        T_{\tau}X^{20}X^{02}
    \right\rangle_{q,\omega_n}
    &=
    \beta\xi^2\delta(q)\delta(\omega_n)
    +
    \left\langle
        T_{\tau}(X^{20}-\xi)(X^{02}-\xi)
    \right\rangle_{q,\omega_n}
    .
%    \label{eqii-01}
\end{align}

Using the above mentioned ``pseudospin'' representation \eqref{eq2-19},
one can express \eqref{eqii-01} in the form
\begin{align}
        \textstyle
    \left\langle
        T_{\tau}(X^{20}-\xi)(X^{02}-\xi)
    \right\rangle_{q,\omega_n}
    &=
    \mathcal{G}^{zz}\sin^2 2\vartheta\,
    +\frac{1}{4}\mathcal{G}^{++}(\cos^2 2\vartheta-1)
    \notag\\
    &\quad{}
    \textstyle
    +\frac{1}{4}\mathcal{G}^{+-}(\cos 2\vartheta+1)^2
    +\frac{1}{4}\mathcal{G}^{-+}(\cos 2\vartheta-1)^2
    +\frac{1}{4}\mathcal{G}^{--}(\cos^2 2\vartheta-1)
    \notag\\
    &\quad{}
    \textstyle
    -\frac{1}{2}(\mathcal{G}^{+z}+\mathcal{G}^{z-})(\cos 2\vartheta+1)\sin 2\vartheta
    \notag\\
    &\quad{}
    \textstyle
    -\frac{1}{2}(\mathcal{G}^{-z}+\mathcal{G}^{z+})(\cos 2\vartheta-1)\sin 2\vartheta
    ,
    \label{eqii-02}
\end{align}
\begin{equation}
    \mathcal{G}^{\alpha\beta}
    =
    \left\langle
        T_{\tau}
        (\sigma^{\alpha}-\langle\sigma^{\alpha}\rangle)_{\tau}
        (\sigma^{\beta}-\langle\sigma^{\beta}\rangle)_{\tau'}
    \right\rangle
    .
    \label{eqii-03}
\end{equation}

Temperature GF in the RPA are derived by means of the Larkin equation
\begin{equation}
    \mathcal{G}^{\alpha\beta}
    =
    \mathcal{G}^{\alpha\beta}_0
    +
    \sum_{\gamma\delta}
    \mathcal{G}^{\alpha\gamma}_0
    j^{\gamma\delta}
    \mathcal{G}^{\delta\beta},
    \label{eqii-04}
\end{equation}
where unperturbed GF are expressed through the single-site GF and the average
values of the pseudospin operator
\begin{align}
    \mathcal{G}^{+-}_0
    &=
    -2 g^0(\omega_n) \langle\sigma_{z}\rangle_0\,,
    &
    \mathcal{G}^{zz}_0
    &=
    \beta b' \delta(\omega_n),
    \notag\\
    \mathcal{G}^{-+}_0
    &=
    -2 g^0(-\omega_n) \langle\sigma_{z}\rangle_0\,,
    &
%    \\
    g^0(\omega_n)
    &=
    \textstyle
    \frac{1}{\text{i} \omega_n + E}\,,
    \label{eqii-05}
    \\
    b(\beta E)
    &\equiv
    \langle\sigma_{z}\rangle_0
    =
    \textstyle
    \frac{1}{2} \tanh \frac{\beta E}{2}\,,
    &
    b'
    &\equiv
    \textstyle
    \frac{\partial b}{\partial (\beta E)}
    =
    \langle\sigma_{z}^2\rangle_0-\langle\sigma_{z}\rangle_0^2\,,
    \notag
\end{align}
and the parameters of the effective interaction look as follows:
\begin{align}
    &\textstyle
    j^{++}(q) = j^{--}(q) = \frac{1}{2}(\cos^2 2\vartheta-1)t_q' \, ,
    \notag\\
    &\textstyle
    j^{+-}(q) = j^{-+}(q) = \frac{1}{2}(\cos^2 2\vartheta+1)t_q' \, ,
    \notag\\
    &
    j^{zz}(q) = 2t_q'\sin^2 2\vartheta \,  ,
    \notag\\
    &
    j^{+z}(q) = j^{z+}(q) = j^{-z}(q) = j^{z-}(q) =
        -t_q' \sin 2\vartheta \cos 2 \vartheta \,  .
    \label{eqii-06}
\end{align}
The equation of the type \eqref{eqii-04} for the ordinary HCB
model was derived in work \cite{Stasyuk09JPS}.

The momentum distribution of particles in the excited state is obtained as
the sum over Matsubara's frequencies
\begin{equation}
    \overline{n}_q - \xi^2\delta(q)
    =
    -\frac{1}{\beta}\sum_{\omega_n} \operatorname{e}^{\text{i}\omega_n 0^{-}}
    \left\langle
        T_{\tau}(X^{20}-\xi)(X^{02}-\xi)
    \right\rangle_{q,\omega_n}
    \label{eqii-07}
\end{equation}
from the temperature GF
\begin{align}
    \left\langle
        T_{\tau}(X^{20}-\xi)(X^{02}-\xi)
    \right\rangle_{q,\omega_n}
    &=
    \langle\sigma^z\rangle_0
    \left(
        \frac{\cos2\vartheta-\Phi_q}
             {\text{i}\omega_n-\varepsilon_q}
        +
        \frac{\cos2\vartheta+\Phi_q}
             {\text{i}\omega_n+\varepsilon_q}
    \right)
   % \notag\\
%    &\quad
    +
    \mathcal{G}_{\text{I}}(q)\cdot\delta(\omega_n),
    \label{eqii-08}
\end{align}
\begin{align}
    \Phi_q
    &=
    \frac{1}{2\varepsilon_q}
    \left[
        E\left(\cos^2\! 2\vartheta+1\right)
        +
        4 \langle\sigma^z\rangle t_q'\cos^2\! 2\vartheta\,
    \right]
    ,
    \label{eq4-17}
    \\
    \varepsilon_q
    &=
    \left[
        \left(E+2\langle\sigma^z\rangle t_q'\right)
        \left(E+2\langle\sigma^z\rangle t_q'\cos^2 2\vartheta\right)
    \right]^{1/2}\!.
    \label{eqii-09}
\end{align}
An explicit form of the spectrum depends on the phase where the system
resides. For example, for NO phase
\begin{equation}
    \varepsilon_q^{\mathrm{(NO)}}=
    \delta-\mu+2\langle\sigma^z\rangle t_q',
    \qquad
    E=\delta-\mu,
    \label{eqii-10}
\end{equation}
while there are two branches $\pm\varepsilon_q^{\mathrm{(SF)}}$ for the SF
phase
\begin{equation}
    \varepsilon_q^{\mathrm{(SF)}} =
    2\abs{\langle\sigma^z\rangle}
    \left\{
        \left(\abs{t_0'}+t_q'\right)
        \left[
            \abs{t_0'}+
            t_q'\frac{(\delta-\mu)^2}{4\abs{t_0'}^2\langle\sigma^z\rangle^2}
        \right]
    \right\}^{1/2},
    \qquad
    E=2\abs{t_0'}\langle\sigma^z\rangle.
    \label{eqii-11}
\end{equation}
This result formally coincides with its analogue for ordinary (two-state) HCB
model (which can be reached in the limit $\delta\to-\infty$). However, an
important difference lies in the fact that the values of
$\langle\sigma^z\rangle$ and $\vartheta$ are defined by an  equation for the
three-state system \eqref{eq2-21}.

\begin{figure}[!h]
\includegraphics[height=0.22\textheight]{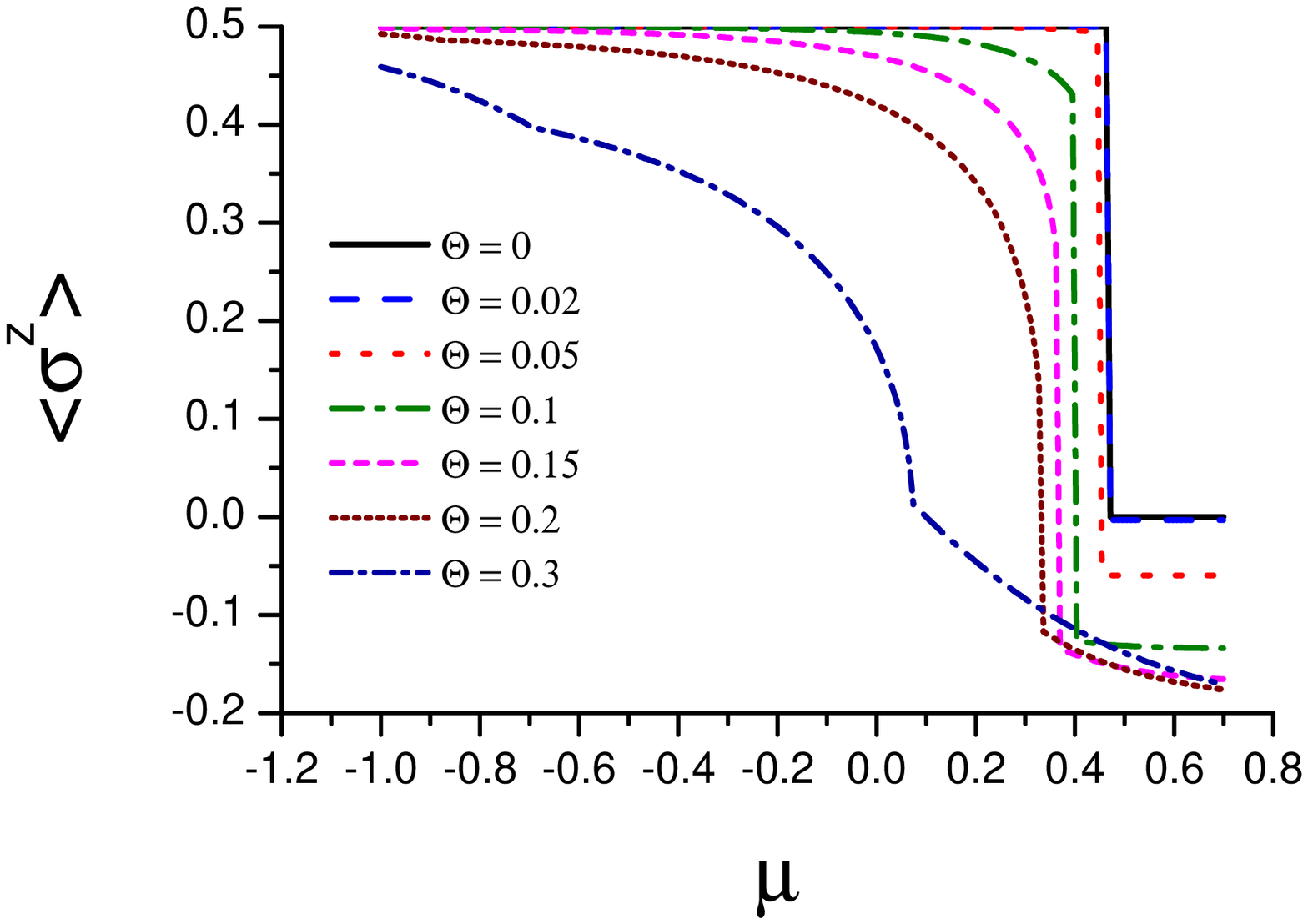}%
\hfill%
\includegraphics[height=0.22\textheight]{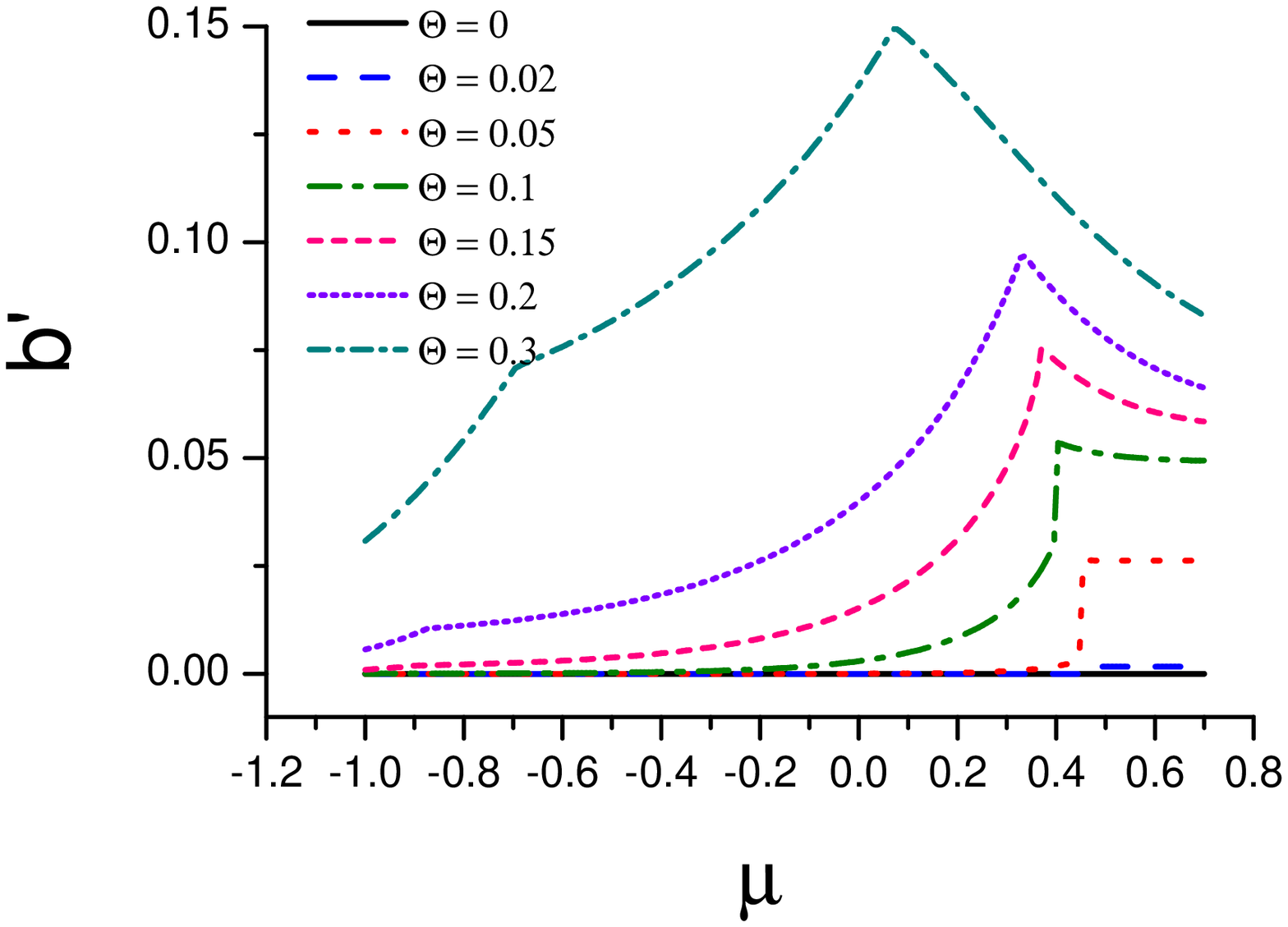}%
\caption{Dependences $\langle\sigma^z\rangle_0$ and $b'$
on chemical potential at different temperatures
($\delta=0.1$, $\abs{t_0'}=1$).}
\label{figb01}
\end{figure}

There are three parts in the momentum distribution:
\begin{equation}
    \overline{n}_q
    =
    \underbrace{%
    \vphantom{
    \left(
        \Phi_q \coth\frac{\beta\varepsilon_q}{2}-\cos2\vartheta
    \right)
    }
    \xi^2\delta(q)}
        _{\text{BE-condensate}}
    +
    \underbrace{%
    \langle\sigma^z\rangle_0
    \left(
        \Phi_q \coth\frac{\beta\varepsilon_q}{2}-\cos2\vartheta
    \right)
    }_{\text{single-particle}\atop\text{excitations}}
    +
        \underbrace{%
        \frac{1}{\beta}\mathcal{G}_{\text{I}}(q)
        }_{\text{non-ergodic}\atop\text{part}}
    .
    \label{eqii-12}
\end{equation}
The first and the second terms correspond to the results obtained in work
\cite{Stasyuk11TMF} while the non-ergodic part (related to the static density
fluctuations) can be obtained only by means of temperature GF:
\begin{align}
    \textstyle
    \!\!\!
    \frac{1}{\beta}
    \mathcal{G}_{\text{I}}(q)
    &=
    \frac{b' \sin^2\! 2 \vartheta}%
    {
        \left(
        \!
        1
        {+}2\frac{\langle\sigma^z\rangle_0}{E} t_q'\cos^2\! 2 \vartheta
        \right)
        \!
        \left(
        \!\!
        1
        {+}2\frac{\langle\sigma^z\rangle_0}{E} t_q' \cos^2\! 2 \vartheta
        {+}2 \beta b' t_q' \sin^2\! 2 \vartheta
        \right)
    },
    \notag\\
    b'
    &= \frac14 \langle \widetilde{X}^{00}+\widetilde{X}^{22} \rangle
        -
        \langle\sigma^z\rangle_0^2\,.
    \label{eqii-13}
\end{align}
The factor $\sin 2 \vartheta$ is non-zero in the SF phase only, so the
existence of the non-ergodic part is also limited to this phase:
\[
    \mathcal{G}_{\text{I}}(q)
    =
    \begin{cases}
        = 0,    &\text{NO phase},\\
        \neq 0, &\text{SF phase}.
    \end{cases}
\]

Due to a sophisticated structure of the nonergodic part, a thorough analysis
of its behavior is rather complicated. Dependencies of $\langle \sigma^z \rangle$ and
$b'$ parameters, entering the formula \eqref{eqii-13}, on chemical potential
and temperature are presented in figure~\ref{figb01}. As one can conclude,
the non-ergodic contribution is ``frozen out'' at low temperatures (because
the hopping is possible between the excited states only).

\begin{figure}[ht]
%\begin{figure}%[p]
%
\centerline{\includegraphics[width=0.49\textwidth]{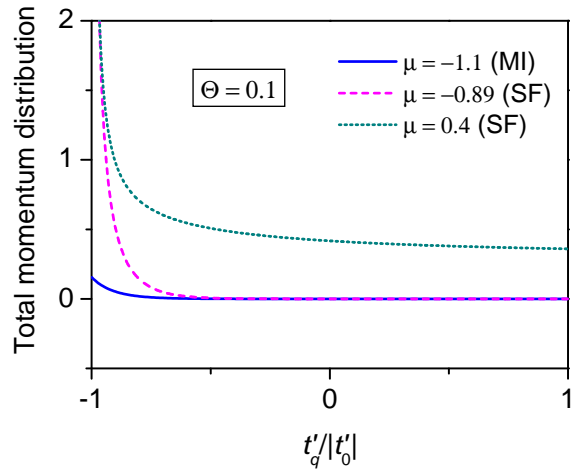}}
\caption{The momentum distribution in the NO and SF phases ($\Theta=0.1$,
$\delta=0.1$, $\abs{t_0'}=1$).}
\label{figb02}
\end{figure}

\begin{figure}%[!b]
\includegraphics[width=0.49\textwidth]{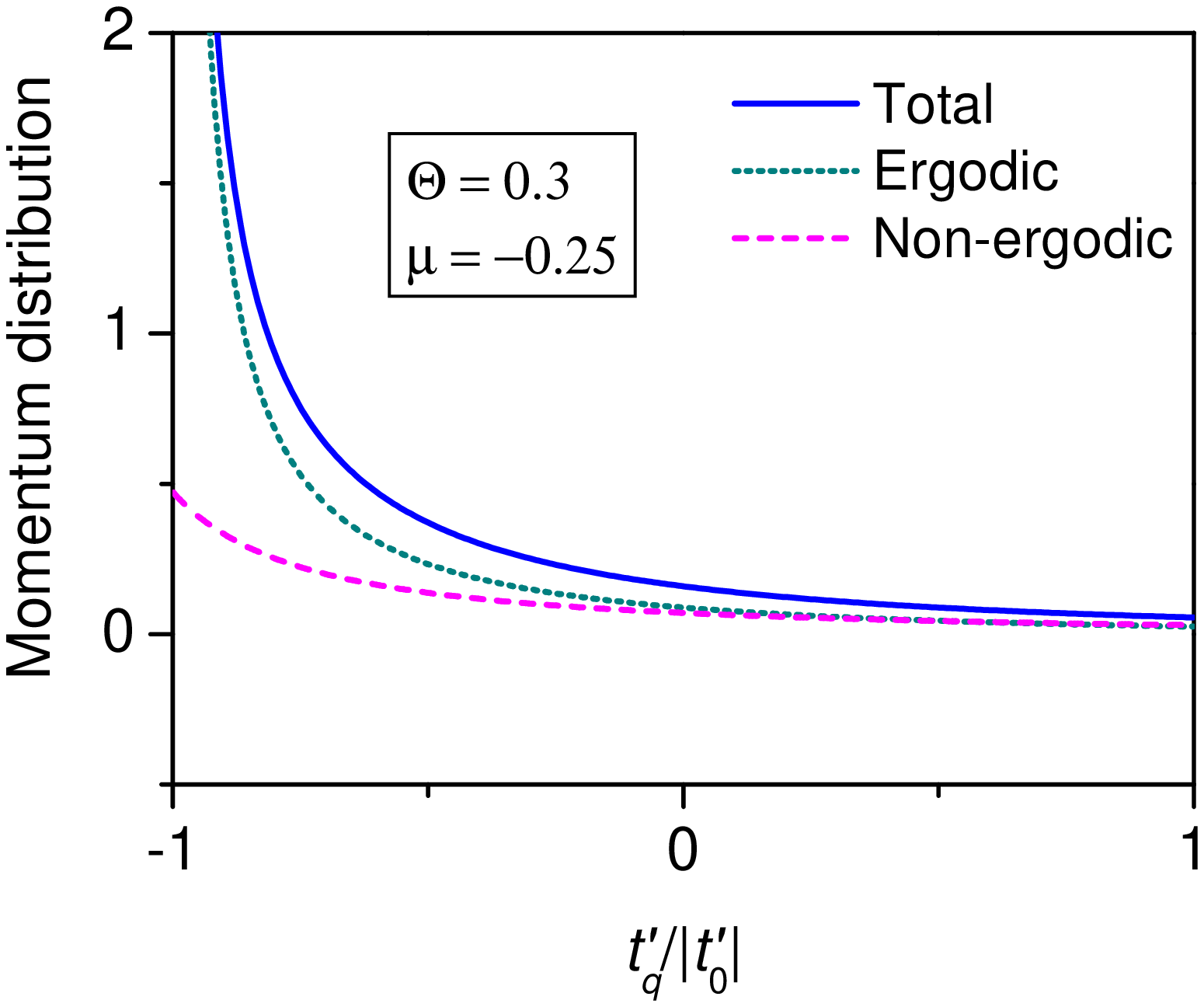}%
\hfill%
\includegraphics[width=0.49\textwidth]{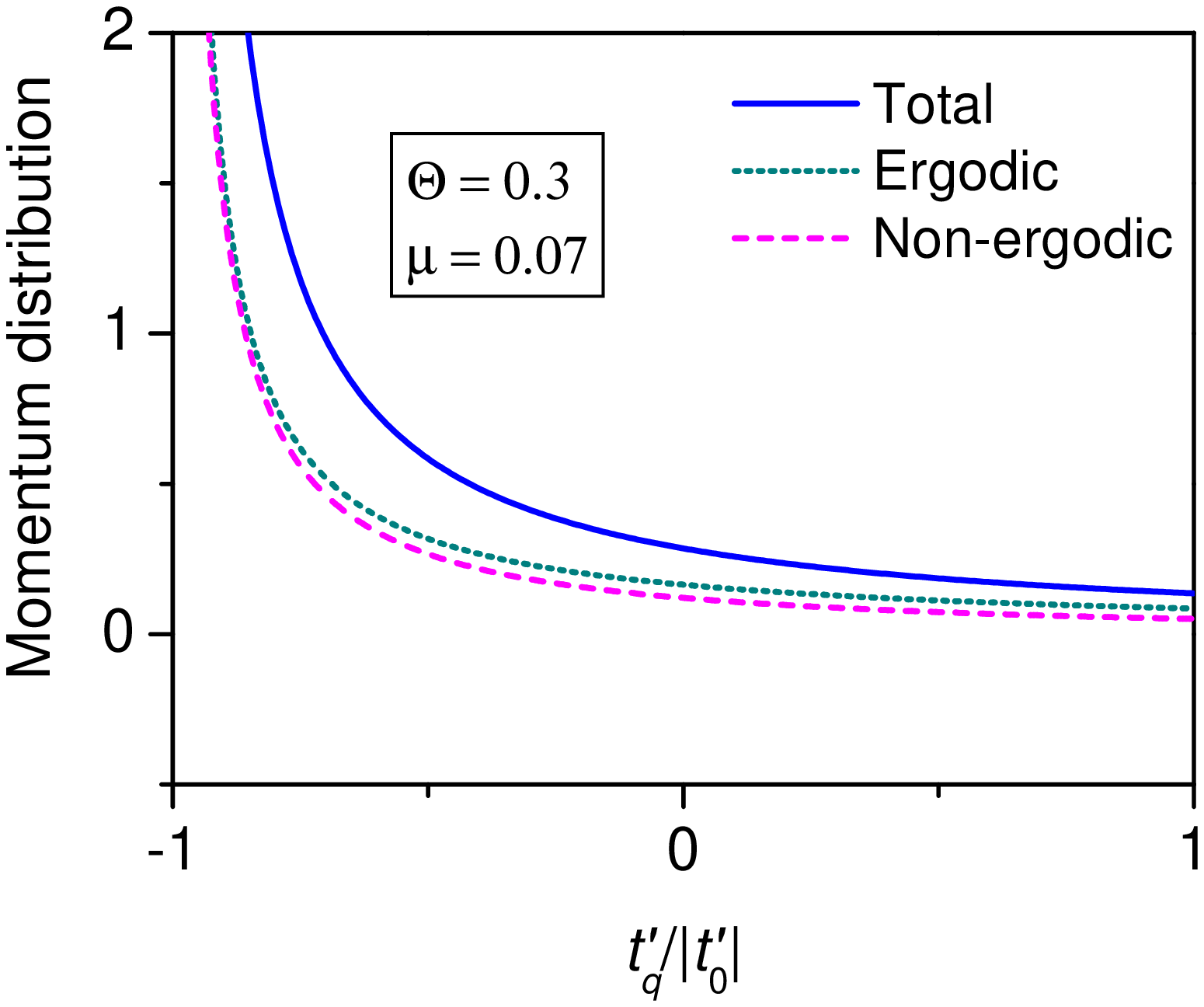}%
\caption{The momentum distribution in the SF phase near
the lines of phase transitions:
comparison of the ergodic and non-ergodic parts
($\Theta=0.3$, $\delta=0.1$, $\abs{t_0'}=1$).}
\label{figb03}
\end{figure}

\begin{figure}[!h]
\vspace{5mm}
\centerline{\includegraphics[width=0.47\textwidth]{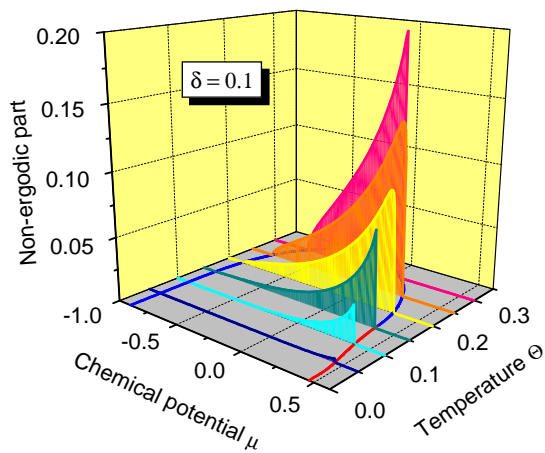}}
\caption{The non-ergodic part $\frac{1}{\beta
N}\sum_q\mathcal{G}_{\text{I}}(q)$ as the function of $\mu$ and $\Theta$
($\delta=0.1$, $\abs{t_0'}=1$).}
\vspace{-5mm}
\label{figb07}
\end{figure}

%As one can see in figure~\ref{figb02},
The momentum distribution, including according \eqref{eqii-12} all three
parts, rapidly increases approaching the centre of the Brillouin zone
(BE-condensation).
The relative role of the non-ergodic part in the total momentum distribution
is illustrated in figures~\ref{figb02}--\ref{figb07}. In the SF phase
$E=2\abs{t_0'}\langle\sigma^z\rangle_0$, therefore


\begin{align}
        \left.
        \left(
        1
        +2\frac{\langle\sigma^z\rangle_0}{E} t_q'\cos^2\! 2 \vartheta
        \right)
        \right\rvert_{q\to0}
        &\to 0
        ,
        \notag \\ \notag
        \frac{1}{\beta} \mathcal{G}_{\text{I}}(q) &\to \infty
\end{align}
near the line of the second order SF-NO transition. Enhancement of the
non-ergodic contribution in the vicinity of the tricritical point is
connected with the increase of the particle number fluctuations (the latter
are characterized by the derivative $\partial n / \partial \mu$, that reduces
to the $b'$ parameter in the simplest approximation). It is confirmed by the
behaviour of the $n(\mu)$ dependence at different temperatures obtained and
illustrated in \cite{Stasyuk11TMF}.

The total occupation of the excited state can be calculated by the summing
over the wave vector (here we substitute the summation by the integration with the
model density of states $g_0(z)$ for a simple cubic lattice
\cite{Stasyuk11TMF})
\begin{equation}
    n_{\text{RPA}}^c
    \equiv
    \overline{n}
    =\frac{1}{N} \sum_q \overline{n}_q
    =\int \overline{n}_z g_0(z)\,\text{d}z.
    \label{eqii-14}
\end{equation}

\begin{figure}[!h]
\includegraphics[width=0.49\textwidth]{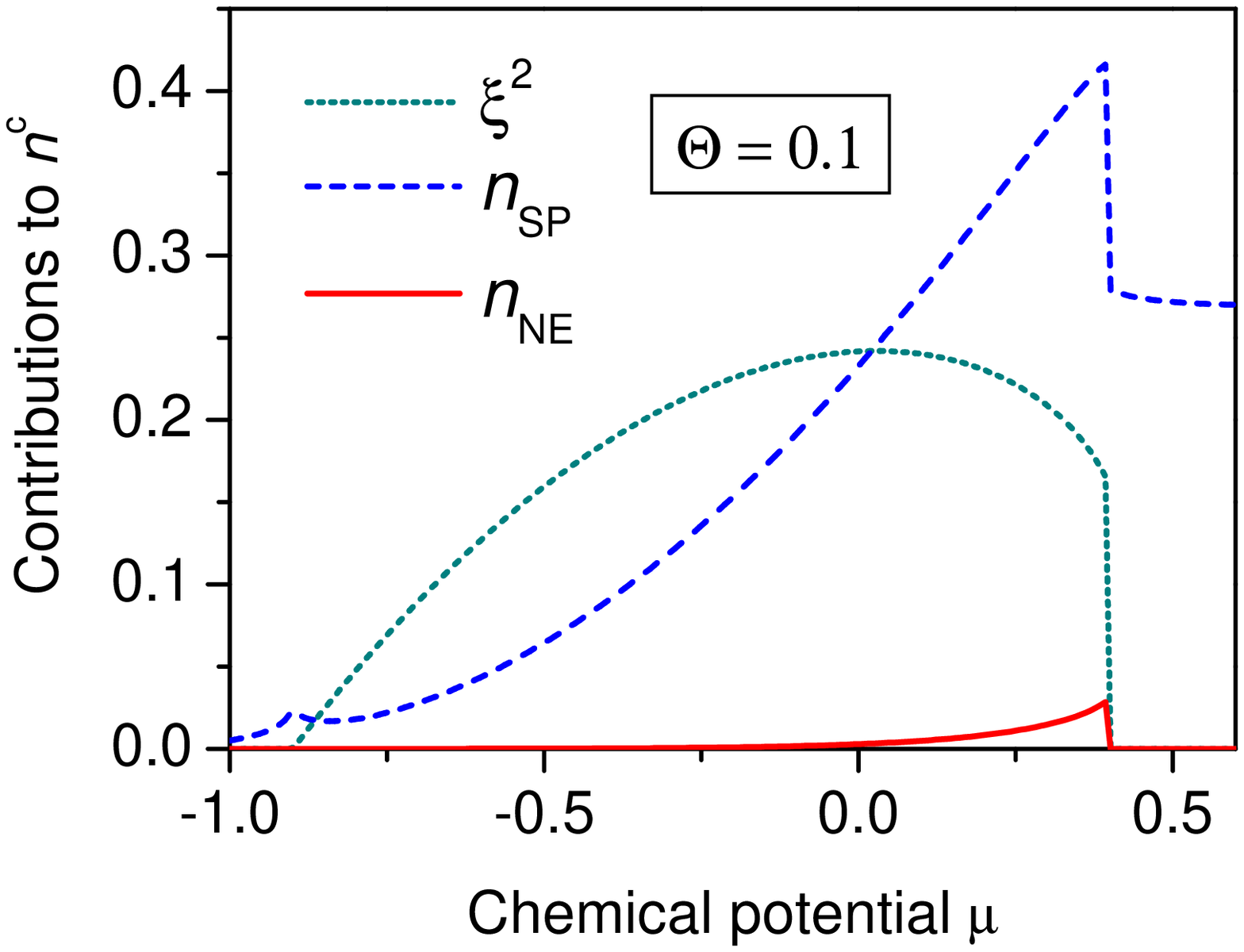}%
\hfill%
\includegraphics[width=0.49\textwidth]{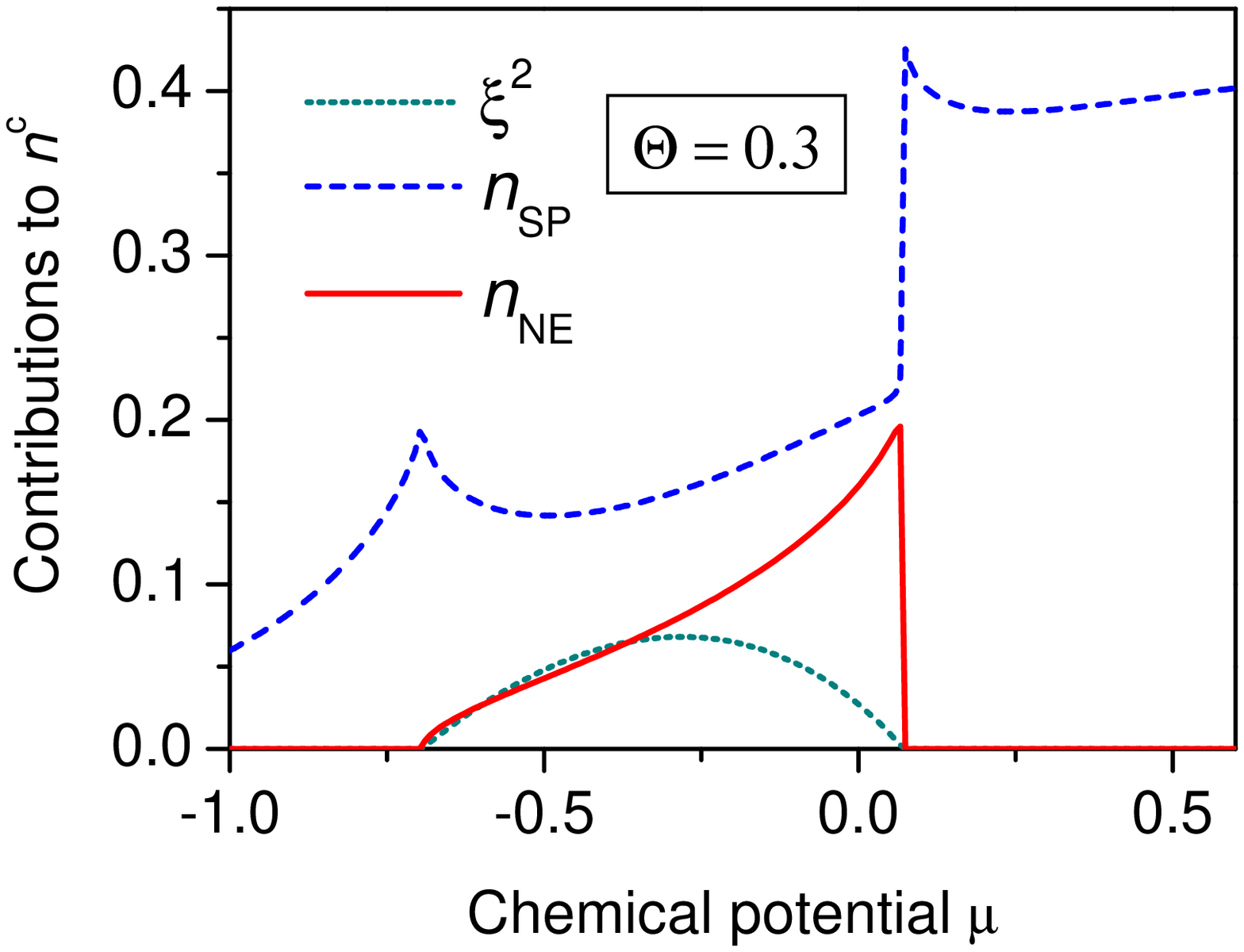}%
\caption{Comparison of the contributions to the excited state:
the BE-condensate, the single-site contribution and the non-ergodic part
($\delta=0.1$, $\abs{t_0'}=1$).}
\label{figb05}
\end{figure}

It is interesting to compare partial contributions of the BE-condensate, the
single-particle excitations and the non-ergodic part in the total occupation
of the excited state:
\[
    n_{\text{RPA}}^c
    \equiv
    \overline{n}
    =\xi^2+n_{\text{SP}}+n_{\text{NE}}\,.
\]
The most pronounced  non-ergodic contribution is visible
near the line of the second order phase transition in the vicinity of the
tricritical point  (figure~\ref{figb05}).

\section{Conclusions}

The non-ergodic part of the momentum distribution (related to the static
density fluctuations) in the excited state for the BHM in the HCB limit is
calculated by means of the temperature GF. This part becomes nonzero only in
the SF phase. Its value rapidly grows and becomes on par with the ergodic
part in the SF phase near the tricritical point (where $\partial{}n /
\partial\mu \to \infty$).

The questions remain, is the non-ergodicity (and the presence of
corresponding contributions to the particle momentum distribution) a
characteristic feature of the hard-core boson limit only and what could be at
the transition to the standard BHM. Such a transition takes place when a
constraint of occupation number $n_i$ is removed and the on-site repulsion
energy $U$ becomes finite. When the non-ergodicity would disappear in this
case, it could mean that the effect is caused by the change of the particle
statistics (due to the transition from Pauli- to Bose-operators). This
problem needs a separate and more detailed consideration.

Dynamical characteristics of the model obtained in the present study and in
the work \cite{Stasyuk11TMF} are generally based on the average values calculated
in the MFA. There is a slight discrepancy between these characteristics and the respective
results in RPA. For a complete self-consistency, one should use the appropriate
expressions for averages, starting with the grand canonical potential
obtained by summation of diagram series consisting closed cycles.

\ukrainianpart

\title{Двостанова модель Бозе-Хаббарда в границі жорстких бозонів:
неергодичність та бозе-конденсація}
\author{І.В. Стасюк, О.В. Величко}
\address{Інститут фізики конденсованих систем НАН України,
79011 Львів, вул. Свєнціцького, 1}

\makeukrtitle

\begin{abstract}
\tolerance=3000%
Досліджується бозе-конденсація в границі жорстких бозонів у моделі
Бозе-Хаббарда з двома локальними станами при переносі бозонів лише у
збудженій зоні. З метою врахування неергодичності одночастинкову спектральну
густину отримано в наближенні хаотичних фаз за допомогою температурних
бозонних функцій Гріна. Неергодичний внесок до функції розподілу частинок за
імпульсом (пов'язаний зі статичними флуктуаціями густини) суттєво наростає і
стає співмірним з ергодичною частиною в надплинній фазі біля трикритичної
точки.
\keywords модель Бозе-Хаббарда, жорсткі бозони, бозе-ейнштейнівська
конденсація, збуджена зона, неергодичність

\end{abstract}

\end{document}